\documentclass[12pt]{article}
\usepackage{amssymb}
\usepackage{euscript}

\let\ssection=\section
\renewcommand{\section}{\setcounter{equation}{0}\ssection}

\newcommand{\be}{\begin{equation}}
\newcommand{\ee}{\end{equation}}

\newcommand{\raw}{\rightarrow}

\newcommand\mathC{\mkern1mu\raise2.2pt\hbox{$\scriptscriptstyle|$}
		{\mkern-7mu\rm C}}				%%% The complex  numbers
		
\newcommand{\om}{\omega} 

 \newcommand{\CB}{{\mathcal B}}
\newcommand{\CC}{{\mathcal C}} \newcommand{\CF}{{\mathcal F}}

\newcommand{\CG}{{\mathcal G}} \renewcommand{\H}{{\mathcal H}}

 \newcommand{\CR}{{\mathcal R}}

\newcommand{\CZ}{{\mathcal Z}}
\def\Dslash{\setbox0=\hbox{$D$}D\hskip-\wd0\hbox to\wd0{\hss\sl/\/\hss}}

\begin{document}
\begin{center}
{\large\bf Emergence and Reduction Combined in Phase Transitions}
\end{center}

\begin{center}
Jeremy Butterfield$^\dagger$ and Nazim Bouatta$^\ddagger$ \vspace{7mm} %%@
\end{center}

\begin{center}
$^\dagger$ {\it Trinity College, Cambridge,  CB2 1TQ, UK}\vspace{5mm}

$^\ddagger$ {\it Department of Applied Mathematics and Theoretical Physics}\\
{\it University of Cambridge, CB3 0WA, UK}
\end{center}

\begin{center} {\tt jb56@cam.ac.uk,  N.Bouatta@damtp.cam.ac.uk} \end{center}
%\vspace{0.6in}

\begin{center} Friday 11 Feb 2011 \end{center}

\begin{center}  For Proceedings of {\em Frontiers of Fundamental Physics, FFP11} \end{center}

\begin{abstract}

In another paper (Butterfield 2011), one of us  argued that emergence and reduction are %%@
compatible, and presented four examples illustrating both. The main purpose of this paper is %%@
to develop this position for the example of phase transitions.

We take it that emergence involves behaviour that is novel compared with what is expected: %%@
often, what is expected from a theory of the system's microscopic constituents. We take %%@
reduction as deduction, aided by appropriate definitions. Then the main idea of our %%@
reconciliation of emergence and reduction is that one makes the deduction after taking a %%@
limit of an appropriate parameter $N$.

Thus our first main claim will be that in some situations, one can deduce a novel behaviour, %%@
by taking a limit $N\to\infty$. Our main illustration of this will be Lee-Yang theory.

But on the other hand, this does not show that the $N=\infty$ limit is ``physically real''. %%@
For our second main claim will be that in such situations, there is a logically weaker, yet %%@
still vivid, novel behaviour that occurs before the limit, i.e. for finite $N$. And it is %%@
this weaker behaviour which is physically real. Our main illustration of this will be the %%@
renormalization group description of cross-over phenomena.

\end{abstract}
\newpage
\tableofcontents

\newpage

%%%%%%%%%%%%%%%%%%%%%%%%%%%%%%%%%%%%%%%%%%%%%
\section{Introduction}\label{Intr}
%%%%%%%%%%%%%%%%%%%%%%%%%%%%%%%%%%%%%%%%%%%%%
This paper develops a position first argued for elsewhere, namely that emergence and %%@
reduction are compatible, despite the
widespread ``ideology'' that they contradict each other (Butterfield 2011, %%@
2011a).\footnote{In physics, this conflict is of course especially associated the debate %%@
surrounding Anderson's famous `More is different' paper (1972). But in this paper we will not %%@
discuss this debate.} Thus Butterfield (2011) presented four examples, each of which %%@
illustrated both emergence
and reduction. One of these examples, treated briefly, was phase transitions; and the main %%@
aim of this paper is to develop this example.

In the rest of this Introduction, we set the stage by defining our terms, and stating our %%@
main claims, about the
reconciliation of emergence and reduction. (For more details about the general issues, we %%@
refer the reader to  Butterfield (2011, Sections 1-3; 2011a).) Then in Section \ref{phtrexpound}, we briefly expound phase transitions, emphasizing: (i) the need for the thermodynamic limit (illustrated mainly by Lee-Yang theory); (ii) critical exponents, universality, scaling and the renormalization group (illustrated mainly by the Ising model). In Section \ref{emergephtrn}, we discuss how the material in Section \ref{phtrexpound} supports our claims: the main illustration will be cross-over phenomena. Section \ref{concl} concludes.

\subsection{Defining terms and stating claims}\label{peace}
`Define your terms!': so goes the refrain. So in order to argue that emergence and reduction %%@
are compatible, we need to first say how we understand these words---which are used in many %%@
ways. It will be clear that our meanings are widespread, rather than heterodox; and more %%@
important, that they are not tailor-made to secure the promised conclusion that they are %%@
compatible! As we shall see, the key to compatibility lies in taking a limit.

We take the ``root-meaning'' of emergence to be behaviour that is novel relative to some %%@
appropriate comparison class. In our examples, as in many discussions, the behaviour in %%@
question is the behaviour of a composite system; and it is novel compared with the behaviour %%@
of the system's components, especially its microscopic or even atomic components. It is also %%@
natural to require that emergent behaviour should be reproducible so that, taken together %%@
with its  connections to other behaviour or properties, it can be systematically investigated %%@
and described. For short, we will call such behaviour `robust'. So for us, emergence is a %%@
matter of novel and robust behaviour.

We take reduction as a relation between theories (of the systems concerned). It is %%@
essentially deduction; though the deduction is usually aided by adding appropriate %%@
definitions linking two theories' vocabularies.  This will be close to endorsing the %%@
traditional philosophical account (Nagel 1961, pp. 351-363; 1979), despite various objections levelled %%@
against it. The broad picture is that the claims of some worse or less detailed (often %%@
earlier) theory can be deduced within a better or more detailed (often later) theory, once we %%@
adjoin to the latter some appropriate definitions of the terms of the former. We also adopt a %%@
mnemonic notation, writing $T_b$ for the better, bottom or basic theory, and $T_t$ for the %%@
tainted, top or tangible theory; (where `tangible' connotes restriction to the observable, %%@
i.e. less detail). So the picture is, with $D$ standing for the definitions: $T_b \& D %%@
\Rightarrow T_t$. In logicians' jargon: $T_t$ is a {\em definitional extension} of $T_b$.

These construals of `emergence' and `reduction' are strong enough to make it worth arguing %%@
that they are in fact consistent. Indeed, they seem to be in tension with each other: since %%@
logic teaches us that valid deduction gives no new ``content'', how can one ever deduce novel %%@
behaviour? (This tension is also shown by the fact that many authors who take emergence to %%@
involve novel behaviour thereby take it to also involve irreducibility.)

The answer to this `how?'  question, i.e. our reconciliation, lies in using limits. One %%@
performs the deduction after taking a limit of some parameter: in such a limit there can be %%@
novelty, compared with what obtains away from the limit. Thus the idea will be that the  %%@
system  is a limit of a sequence of systems, typically as some parameter (in the theory of %%@
the systems) goes to infinity (or some other crucial value, often zero); and its  behaviour %%@
is novel and robust compared to those of  systems described with a finite (respectively: %%@
non-zero) parameter.

Butterfield (2011) described four examples of this. Each example is a model, or a framework %%@
for modelling, from well-established mathematics or physics; and each involves an integer %%@
parameter $N = 1, 2, ...$ and its limit $N \raw \infty$. In three of the examples, $N$ is  %%@
the number of degrees of freedom of the system, or a  very similar notion; and the novel %%@
behaviour at the limit depends on new
mathematical structures that arise for infinite systems. (In the fourth example, there was no %%@
infinite system; and so the limiting process must be expressed in terms of quantities and %%@
their values, rather than systems.)

One of these examples was this paper's topic: phase transitions. Here, the limit is the %%@
thermodynamic limit: i.e. roughly, letting the number
$N$ of constituent particles tend to infinity, while other parameters stay constant or scale %%@
appropriately (e.g. the density is constant, so that the volume also goes to infinity). Thus %%@
our first main claim about phase transitions is that deductions of novel and robust behaviour %%@
in this limit yield examples combining emergence and reduction. (Compare (1:Deduce) below for %%@
a formal statement.) We will illustrate this with KMS states, and
(in greater detail) Lee-Yang theory (Sections \ref{needTD} and \ref{phtrnwithwithoutpaul}).

In the recent philosophical literature, some
authors have alleged that the thermodynamic limit, and similar $N = \infty$ limits, are %%@
``physically real'' in a strong sense. We deny this, but we will not go into the details here %%@
(for which, cf. Butterfield 2011, especially Section 3). After all, one might say that our denial is hardly news, %%@
since in the
case of the thermodynamic limit, denying that the number $N$ of constituent particles is %%@
finite is apparently tantamount to denying atomism!

But we agree---and Butterfield (2011) exhibited in his
examples---that there is a logically weaker, yet still vivid, novel and robust behaviour that %%@
occurs {\em before} one gets to the limit. Thus in the phase transitions example: for finite %%@
$N$. And, we maintain, it is this weaker behaviour which {\em is} physically real. In this %%@
paper, we will illustrate this with the Ising model, and {\em cross-over phenomena} for critical phase transitions (Sections \ref{Ising} and \ref{crossover}). Understanding such transitions, and in particular
cross-over, is one of the great successes of  renormalization group ideas. In
order to prepare for Section \ref{crossover}, we will in Sections \ref{critexp} and \ref{WRG} summarize
those ideas.

So to sum up: there is a contrast between a strong sense of emergence, which is absent away %%@
from the limit (absent at finite $N$); and a weak sense of %%@
emergence which is present away from the limit (at finite $N$). And these two senses yield %%@
our two main claims. The first is:
\begin{quote}
 (1:Deduce): Emergence is compatible with reduction. And this is so, with a strong %%@
understanding both of `emergence' (i.e. `novel and robust behaviour') and of `reduction' %%@
(viz. logicians' notion of definitional extension). For considering a limit $N \raw \infty$ %%@
enables one to {\em deduce} novel and robust behaviour, in strong senses of `novel' and %%@
`robust'.\footnote{Butterfield (2011) discusses how
choosing a weaker theory using finite $N$ blocks the deduction of this strong sense. Since %%@
the theories $T_t$ and $T_b$ are often defined only  vaguely (by labels like `thermodynamics' %%@
and `statistical mechanics'), this swings-and-roundabouts situation explains away some of the %%@
frustrating
controversies over ``reductionism'', i.e. over whether $T_t$ is reducible to $T_b$.} Our main %%@
illustration of this will be Lee-Yang theory, expounded in Section \ref{needTD}.
\end{quote}
The second claim is:
\begin{quote}
\indent (2:Before): But on the other hand: emergence, in a logically weaker yet still vivid %%@
sense, occurs {\em before} we get to the limit. That is: one can construe `novel and robust %%@
behaviour' weakly enough that it {\em does} occur for finite $N$. Our main illustration of %%@
this will be the renormalization group description of cross-over. We sketch the renormalization group in
Sections \ref{critexp} and \ref{WRG}; and describe cross-over in  Section \ref{crossover}.
\end{quote}

\section{Phase transitions}\label{phtrexpound}
We begin by defining our topic more closely (Section \ref{seplimiting}). Then we sketch how statistical mechanics treats phase transitions by taking a limit, in
which the number of constituent particles (or sites in a lattice) $N$ goes to infinity %%@
(Section \ref{needTD}). Then we introduce critical phenomena (Section \ref{approach}) and the renormalization group (Section \ref{WRG}). As we proceed, it will be clear  which material in this Section supports our claims (1:Deduce) and (2:Before).

\subsection{Separating issues and limiting scope}\label{seplimiting}

Phase transitions form one aspect of a very large topic, the ``emergence'' of thermodynamics %%@
from
statistical mechanics---around which debates about the reducibility of one to the other %%@
continue. This topic is very large, for various reasons. Three obvious ones are: (i) both %%@
thermodynamics and statistical mechanics are entire sciences, rather than single theories of %%@
manageable size (as Section \ref{Intr}'s mnemonic notation $T_b, T_t$ suggested); (ii) %%@
statistical mechanics, and
so this topic, can be developed in either classical or quantum terms; (iii) there is no
single agreed formalism for statistical mechanics (unlike e.g. quantum mechanics).

Phase transitions are themselves a large topic: there are several classification schemes for %%@
them, and various approaches to understanding them---some of which come in both classical and %%@
quantum versions. We will specialize to one crucial aspect; which will however be enough to %%@
illustrate the main claims. Namely: the fact (on most approaches!) that for statistical %%@
mechanical systems, getting a (theoretical description of) a phase transition requires that %%@
one take
`the' thermodynamic limit, in which the number of constituent particles (or sites in a %%@
lattice) $N$ goes to infinity. We say `the', since the details vary from case to case. More %%@
details in Section \ref{needTD}; but the idea is that both the number $N$  of constituent %%@
particles (or sites), and the volume $V$ of the system  tend to infinity, while the density
$\rho = N/V$ remains fixed.

Even for this one aspect, we will restrict our discussion severely. Three main limitations %%@
are:\\
\indent 1): As regards philosophy, we impose a self-denying ordinance about recent %%@
controversies whether the thermodynamic limit ($N\to\infty$) is ``physically real'', and %%@
whether a ``singular'' limit is necessary for emergence. (Butterfield (2011) discusses these: %%@
citing e.g. Callender (2001, Section 5, pp.
547-551), Liu (2001, Sections 2-3, pp. S326-S341), Batterman (2005, Section 4, pp. 233-237);  %%@
and more recently, Mainwood (2006, Chapters 3,4; 2006a) and Bangu (2009, Section 5, pp. %%@
496-502)).\\
\indent 2): In physics, how to understand phase transitions is an ongoing research area. For %%@
our purposes, the main limiting (i.e. embarrassing!) fact is that most systems do {\em not} %%@
have a well-defined thermodynamic limit---so that all that follows is of limited scope.\\
\indent 3): Various detailed justifications can be given for phase transitions requiring us %%@
to take the thermodynamic limit; and Section \ref{needTD} will only sketch a general %%@
argument, and mention two examples. Much needs to be (and has been!) said by way of assessing %%@
these justifications---but we will not enter into this here. But by way of emphasizing how %%@
open all these issues are, we note that some physicists have developed frameworks for %%@
understanding phase transitions {\em without} taking the thermodynamic limit (Gross (2001)).

Finally, by way of limiting our scope: we mention that for further information, we
recommend, in addition to the famous monographs, such as Ruelle (1969): \\
\indent (i)  accounts by masters of the subject, such as Emch and Liu (2002, Chapters 11-14) and
Kadanoff (2009, 2010, 2010a), which treat the technicalities and history, as well as the
conceptual foundations, of the subject; and \\
\indent (ii) Mainwood (2006, Chapters 3,4; 2006a), to which we are indebted, especially in Section %%@
\ref{phtrnwithwithoutpaul}'s treatment of phase transitions in finite $N$ systems, and %%@
Section \ref{crossover}'s discussion of cross-over.

%%%%%%%%%%%%%%%%%%%%%%%%%%%%%%%%%%%%%%%%%%%%%%%
\subsection{The need for the thermodynamic limit}\label{needTD}
%%%%%%%%%%%%%%%%%%%%%%%%%%%%%%%%%%%%%%%%%%%%%%%
We will first give a broad description of the need for the thermodynamic limit (Section %%@
\ref{singies}).  Then in Section \ref{YLKMS}, we give two
 examples showing how the limit secures new mathematical structure
appropriate for describing phase transitions: Lee-Yang theory and KMS states.

\subsubsection{Singularities}\label{singies}
For classical physics, the brutal summary of why we
need the thermodynamic limit is as follows. Statistical mechanics follows thermodynamics in %%@
representing
phase transitions by non-analyticities of the free energy $\CF$. But a non-analyticity cannot %%@
occur for the free energy of a system with finitely many constituent particles (or %%@
analogously: lattice sites). So statistical mechanics considers a system with infinitely many %%@
particles or sites, $N = \infty$. One gets some control over this idea by subjecting the %%@
limiting process, $N \raw \infty$, to physically-motivated conditions like keeping the %%@
density constant, i.e. letting the volume $V$ of the system also go to infinity, while $N/V$ %%@
is constant. This infinite
limit gives new mathematical structures: which happily turn out to describe phase %%@
transitions---in many cases, in remarkable quantitative detail.

We spell out this line of argument in a bit more detail. In Gibbsian statistical mechanics, %%@
we postulate that the probability of a state $s$ is proportional to $\exp(-\H(s)/k_BT) \equiv %%@
\exp(-\beta \H(s))$, where $\H$ is the Hamiltonian,  $\beta := 1/k_BT$ is the inverse %%@
temperature and $k_B =: k$ is Boltzmann's constant. That is:
\be
{\rm{prob}}(s) = \exp(-\beta\, \H(s)) / \CZ
\label{classlGibbs}
\ee
where the normalization factor $\CZ$, the partition function, is the sum (or integral) over %%@
all states, and defines the free energy $\CF$ as:
\be
\CZ \; = \; \sum_s \exp(-\beta\,\H(s)) \; =: \; \exp(-  \beta\, \CF) \; .
\label{defZF}
\ee
Thus $\CF$ is essentially, the logarithm of the partition function; which is itself the sum %%@
(or integral) over all states of the exponential of the Hamiltonian. It turns out that $\CZ$ %%@
and $\CF$ encode, in their functional forms, a great deal of information about the system: %%@
various quantities, in particular the system's thermodynamic quantities, can be obtained from %%@
them, especially by taking their derivatives. For example, in a ferromagnet, the %%@
magnetization is the first derivative of the free energy with respect to the applied magnetic %%@
field, and the magnetic susceptibility is the second derivative.

Now, broadly speaking: phase transitions involve abrupt changes, in time and-or space, in %%@
thermodynamic quantities: for example, think of the change of particle density in a %%@
solid-liquid, or liquid-gas, transition. Thermodynamics describes these changes as %%@
discontinuities in thermodynamic quantities (or their derivatives), and statistical mechanics %%@
follows  suit. This means that the statistical mechanical description of phase transitions %%@
requires non-analyticities of the free energy $\CF$. But under widely applicable assumptions, %%@
the free energy of a system with finitely many constituent particles (or analogously: sites) %%@
is an analytic function of the thermodynamic quantities within it. For example, in the Ising %%@
model with $N$ sites, the Hamiltonian $\H$ is a quadratic polynomial in spin variables (cf. %%@
Eq.(\ref{IsiHamn})). This means that the partition function $\CZ$, which by Eq.(\ref{defZF}) %%@
is a sum of exponentials of $- \beta\, \H$, is analytic; and so also is its logarithm, and %%@
the free energy. (This and similar arguments about more general forms of the Hamiltonian (or %%@
partition function or free energy), are widespread: e.g. Ruelle (1969, p. 108f.), Thompson %%@
(1972, p. 79), Le Bellac (1991, p. 9), Lavis and Bell (1999, pp. 72-3)) .

We of course admit that---as the phrases `broadly speaking' and `under widely applicable %%@
assumptions' indicate---this argument why phase transitions need the thermodynamic limit is %%@
not a rigorous theorem. Hence the efforts mentioned in Section \ref{seplimiting} to develop a theory of phase
transitions in finite systems; and the philosophical debate among Callender and others about the reality and  singularity of limits. Hence also the historical struggle to recognize the need for %%@
infinite systems: both Emch and Liu (2002, p. 394) and Kadanoff (2009, p. 782; 2010, Section %%@
4.4) cite the famous incident of Kramers putting the matter to a vote at a meeting in memory %%@
of Van der Waals in 1937.\footnote{Of course, since we have not precisely defined %%@
`thermodynamic limit'---let alone `phase transition'!---the argument could hardly be a %%@
rigorous theorem. Ruelle (1969, Sections 2.3-4, 3.3-5) rigorously discusses conditions  for %%@
the thermodynamic limit; (cf. also Emch (1972, p. 299; 2006, p. 1159) Lavis and Bell (1999a, %%@
pp. 116, 260)). Such discussions bring out how in some models, the  limit is not just the %%@
idea that, keeping the density constant, the number $N$ of molecules or sites tends to %%@
infinity: there are also conditions on the limiting behaviour of short-range forces. This is also illustrated at the start of Section \ref{YLKMS}.}

 But this argument, although not a rigorous theorem, is very ``robust''---and recognized as %%@
such by the literature. For example, Kadanoff makes it one of the main themes of his recent %%@
discussions, and even  dubs it the `extended singularity theorem' (2010, Sections 2.2, 6.7.1; %%@
2010a, Section 4.1). He also makes it a playful variation on Anderson's slogan that `more is %%@
different'. Namely, he summarizes it in Section titles like `more is the same; infinitely %%@
more is different' (2009, Section 1.5; 2010, Section 3). In any case, for the rest of this %%@
paper, we accept the argument.

\subsubsection{New mathematical structures: Lee-Yang theory, and KMS
states}\label{YLKMS}
Taking the thermodynamic limit introduces new mathematical structures. But the variety of %%@
formalisms in statistical mechanics (and
indeed, the variety of justifications for taking the limit) means that there is a concomitant %%@
variety of new structures that in the limit get revealed. We describe one classical example, %%@
Lee-Yang theory; and one quantum example, KMS states. It will be obvious how this illustrates our claim (1:Deduce).

In Lee-Yang theory (initiated by Yang and Lee 1952), one uses complex generalizations of the %%@
partition function and free energy, writing $\CZ(z)$ and $\CF(z)$, and then argues that for any $z \in %%@
\mathC$, there can be a phase transition (i.e. a non-analyticity of $\CF$ or $\CZ$) at $z$ only %%@
if there are zeroes of $\CZ$ arbitrarily close to $z$. For finite $N$, $\CZ$ has finitely many %%@
zeroes, so that there can be a phase transition only at the zeroes themselves: but all of %%@
them lie off the real line, and so are not physical. Taking the limit $N \raw \infty$ %%@
``breaks'' this last argument: there can be a curve of zeroes that intersects the real axis. %%@
Indeed, in Lee-Yang theory one goes on to classify phase transitions in terms of the %%@
behaviour of the density of zeroes in $\mathC$: (cf. Thompson 1972, pp. 85-88; Ruelle 1969, %%@
pp. 108-112; Lavis and Bell 1999a, pp. 114, 125-134). Here, we will follow the pedagogical account  given by Blythe and Evans (2003).

We can illustrate the Lee-Yang theory with their original paper's model. Consider a system of up to $N$ classical spherical particles in a box of finite volume $V$; (so the upper bound $N$ arises from the particles' finite size). The interaction of these particles is described by a potential $V(r_{ij})$, where $r_{ij}$ is the distance between the $i$th and the $j$th particles.
We assume:
\begin{itemize}
\item  The particles have an impenetrable core of diameter $a$ : $V(r_{ij}) = \infty$ if $r_{ij} < a$. 
\item The interaction has a finite range $b$ : $V(r_{ij}) =0$ if $b< r_{ij}$.
\item $V(r_{ij}) = -{\epsilon_{ij}}$ if $a\leq r_{ij}\leq b$.
\end{itemize}

If the system  interacts with an external heat reservoir at a given chemical potential $\tilde{\mu}$ per particle, the partition function is then given, as a function of $z := e^{\beta\tilde\mu}$, by

\be
\CZ_N(z) = \sum_{n=0}^N \frac{z^n}{\lambda^{3n}_T\,n!}Q_n\, , 
\ee
where $\lambda_T$ is the thermal wavelength and $Q_n$ is the configuration integral, the integral over the configuration coordinates: 

\be
 Q_n := \int d\rm{q}^n \exp\Big(-\beta\sum_{i<j = 1}^{n(n-1)/2} V(r_{ij})\Big)\, .
\ee

One can now analyse the zeroes of the partition function by using the variable %%@
$z$, and one expresses the partition function $\CZ_N$ as a %%@
polynomial of degree $N$ in terms of $z$

\be
\CZ_N(z)  \propto \prod_{n=1}^N\left(1-\frac{z}{z_n}\right)\, %%@
,\label{partfuncfact}
\ee
where the  $z_n$ are  the $N$ zeroes of the partition function. For finite $N$, the partition function $\CZ$ %%@
will have finitely many zeroes, and all of them lie in the complex plane off the real line.

To explore in detail the locations of the zeroes of $\CZ$, we consider the {\em complex} %%@
generalization of the free energy

\be
h_N(z) \equiv \frac{\ln \CZ_N(z)}{N} = %%@
\frac{1}{N}\sum_{n=1}^N\ln\left(1-\frac{z}{z_n}\right)\, .
\ee
One may differentiate $h_N(z)$ infinitely many times over any region of the complex plane %%@
that does not contain any zeroes: for the Taylor expansion of $h_N(z)$ around a point $z\neq %%@
z_n$ has a finite radius of convergence, given by the distance from $z$ of the nearest zero.

Now, in order to explore the possible phase transitions in Lee-Yang theory, we turn to the %%@
thermodynamic limit $N\to\infty$ and consider the behaviour of the free energy $h_N(z)$ in %%@
this limit:

\be
h(z) := \lim_{N\to\infty} \frac{Z_N(z)}{N} \equiv \int dz' %%@
\rho(z')\ln\left(1-\frac{z}{z'}\right)\, , \label{density}
\ee
where $\rho(z')$ is the local density of zeroes in the complex plane. To analyse the zeroes %%@
of the expression Eq.(\ref{density}), one may introduce a {\em potential} $\varphi(z)\equiv %%@
\rm{Re} \;$$h(z)$. If one takes the Laplacian of the right and left sides of the real part of %%@
Eq.(\ref{density}), one may express the local density $\rho(z)$ in terms of the potential %%@
$\varphi(z)$
\be
\rho(z) = \frac{1}{2\pi}\nabla^2\varphi(z)\, .\label{electrostatic}
\ee
Eq.(\ref{electrostatic}) is similar to Poisson's equation, familiar from electrostatics;
where $\rho(z)$ and $\varphi(z)$ are analogous to the charge density and to the electrostatic %%@
potential respectively. Using the electrostatic analogy, the relevant solutions for the %%@
potential $\varphi$ may be described by a curve that intersects the positive real axis.

To sum up: in the thermodynamic limit, the existence of a curve of zeroes
intersecting the real axis leads   to  non-analyticities in the partition function and
therefore to physical phase transitions.

We turn to our quantum example: KMS states. The leading ideas are that these are a %%@
generalization, for systems with infinitely many degrees of freedom, of Gibbs states; and %%@
that considering an infinite system allows for unitarily
inequivalent representations of the algebra of quantities, with these representations %%@
differing in the value of a global, or macroscopic, quantity---as is needed to describe phase %%@
transitions. (For more details, cf. e.g. Emch (1972, pp. 213-223; 2006, Section 5.6-7, pp. %%@
1144-1154); Sewell (1986, pp. 73-80; 2002, pp. 113-123); Emch and Liu (2002, pp. 346-357),  %%@
Liu and Emch (2005, pp. 142-145, 157-161).)

Thus we recall that the Gibbs state of a finite quantum system with Hamiltonian $\H$ at %%@
inverse temperature $\beta = \frac{1}{kT}$ is given by the density matrix
\begin{equation}
\rho = \exp(- \beta\,\H) / {\rm tr}(\exp(- \beta\,\H))
\end{equation}
and represents the (Gibbsian) equilibrium state of the system. Note the beautiful analogy %%@
with Eq.s (\ref{classlGibbs}) and (\ref{defZF})! But it is unique (for given $\beta$): %%@
thereby precluding the representation of two phases of the system at a common %%@
temperature---as one would want for a phase transition.\footnote{This uniqueness also %%@
precludes spontaneous symmetry breaking, understood (as usual) as the allowance of distinct %%@
equilibria that differ by a dynamical symmetry. Spontaneous symmetry breaking is (yet %%@
another!) important aspect of phase transitions: a fine recent philosophical discussion is %%@
Liu and Emch (2005).}

So how can we give a quantum description of phase transitions? The algebraic approach to %%@
quantum statistical mechanics proposes some states, viz. KMS states, which are defined on %%@
infinite quantum systems  and which generalize the notion of a Gibbs state in a way that is %%@
(a) compelling mathematically, and (b) well-suited to describing phase transitions. A word %%@
about each of (a) and (b):---\\
\indent (a): A mathematical property of Gibbs states (the `KMS condition') is made into a %%@
definition of an equilibrium state that is applicable to both infinite and finite systems: %%@
(for the latter it coincides with the Gibbs state at the given temperature). KMS states can %%@
be shown to have various stability or robustness properties that makes them very well suited %%@
to describe (stable) physical equilibria. (Emch (2006, Section 5.4, pp. 1128-1142) is a %%@
excellent survey of these properties. Such a survey brings out how KMS states are
themselves another example of emergent behaviour, in our sense of novel and robust %%@
properties!)   \\
\indent (b): The set $K_{\beta}$ of KMS states at a given inverse temperature $\beta$ is in %%@
general not a singleton set. Rather, it is convex, with: (i) every element having a unique %%@
expression as a mixture of its extremal points; and (ii) its extremal points being %%@
well-suited to describe pure thermodynamical phases (mathematically, they are factor states). %%@
Taken together, (i) and (ii) suggest that a compelling representative of the state of a %%@
system undergoing a phase transition at inverse temperature $\beta$ is a non-extremal $\om %%@
\in K_{\beta}$.

So much by way of indicating the need and justification for taking ``the'' thermodynamic
limit. We turn to discussing {\em the approach} to limits: both the thermodynamic limit, and others.

%%%%%%%%%%%%%%%%%%%%%%%%%%%%%%%%%%%%%%%%%%%%%
\subsection{Approaching limits}\label{approach}
%%%%%%%%%%%%%%%%%%%%%%%%%%%%%%%%%%%%%%%%%%%%
We first discuss the approach to the thermodynamic limit, for the specific case of the Ising %%@
model (Section \ref{Ising}). It will be obvious how this illustrates our claim (2:Before). This will lead in to Section \ref{critexp}'s discussion of critical exponents, %%@
universality and scaling---which presages the renormalization group in Section \ref{WRG}. Sections \ref{critexp} and \ref{WRG} will thus be mostly concerned with approaching the critical point, rather than the thermodynamic limit.

\subsubsection{The Ising model introduced}\label{Ising}
A paradigm classical example of approaching the limit $N \to \infty$ of infinitely many sites %%@
in a lattice is the change of magnetization, at sub-critical temperatures, of a ferromagnet: %%@
as described by the Ising model (in two or more
spatial dimensions).

The Ising model postulates that at each of $N$ sites, a classical ``spin'' variable $\sigma$ %%@
(which we think of as defined with respect to some spatial direction) takes the values $\pm %%@
1$. To do Gibbsian statistical mechanics, i.e. to apply Eq.s (\ref{classlGibbs}) and %%@
(\ref{defZF}), we need to define a Hamiltonian and then sum over configurations. The %%@
Hamiltonian is chosen to give a simple representation of the ideas that (i) neighbouring %%@
spins interact and tend to be aligned (i.e. their having equal values has lower energy) and %%@
(ii) the spins are coupled to an external magnetic field which points along the given spatial %%@
direction. Thus the Hamiltonian is
\be
\H \; \; =  - \; \; J \; \sum_{p,q} \sigma_p \sigma_q \;\; +\; \; \mu B \; \sum_{p}\sigma_p %%@
\; ,
\label{IsiHamn}
\ee
where: the first sum is over all pairs of nearest-neighbour sites, the second sum is over all %%@
sites, $J$ being positive represents the fact that the neighbouring spins ``like'' to be %%@
aligned, $\mu$ is a magnetic moment and $B$ is an external magnetic field.

The simplest possible case is the case of $N = 1$! With only one site, the Hamiltonian  %%@
becomes
\be
\H = \mu B\; \sigma  \; ;
\label{IsiOneSpin}
\ee
so that if we define a dimensionless coupling $H := -\mu B/kT$, then Eq.s (\ref{classlGibbs}) %%@
and (\ref{defZF}) give
\be
{\rm{prob}}(+1) = e^H / z \;\; {\rm{and}} \;\; {\rm{prob}}(-1) = e^{-H} / z \;\; , \; %%@
{\rm{with}} \;\;
z = e^{H} + e^{-H} = 2 \cosh H \; .
\label{Isi1}
\ee
This implies that the magnetization, i.e. the average value of the spin, is
\be
\langle \sigma \rangle = e^{H}/z - e^{-H}/z = \tanh H \; .
\ee
This is as we would hope: the statistical mechanical treatment of a single spin predicts that the %%@
magnetization increases smoothly from -1, through zero, to +1 as the applied field along the %%@
given axis increases from minus infinity through zero to plus infinity.

We can ask (as we did for Lee-Yang theory): what about larger $N$? The analytical problem %%@
becomes much more complicated (though the
magnetization is still a smooth function of the applied field). But the effect is what we %%@
would expect: a larger $N$ acts as a brake on the ferromagnet's response to the applied field %%@
increasing from negative to positive values (along the given axis).  That is: the increased %%@
number of nearest neighbours means that the ferromagnet ``lingers longer'', has ``more %%@
inertia'', before the rising value of the applied field succeeds in flipping the %%@
magnetization from -1 to +1. More precisely: as $N$ increases, most of the change in the %%@
magnetization occurs more and more steeply, i.e. occurs in a smaller and smaller interval %%@
around the applied field being zero.  Thus the magnetic susceptibility, defined as the %%@
derivative of magnetization with respect to magnetic field, is, in the neighbourhood of 0, %%@
larger for larger $N$, and tends to infinity as $N \rightarrow \infty$. As Kadanoff says: `at %%@
a very large number of sites ... the transition will become so steep that the causal observer %%@
might say that it has occurred suddenly. The astute observer will look more closely, see that %%@
there is a very steep rise, and perhaps conclude that the discontinuous jump occurs only in %%@
the infinite system' (2009, p. 783; and Figure 4).

This general picture of the approach to the $N \raw \infty$ limit
applies much more widely. In particular, very similar remarks apply to liquid-gas phase %%@
transition, i.e. boiling. There the quantity which becomes infinite in the $N \rightarrow %%@
\infty$ limit, i.e. the analogue of the magnetic susceptibility, is the compressibility, %%@
defined as the derivative of the density with respect to the pressure.

\subsubsection{Critical exponents, universality and scaling}\label{critexp}
We now turn to the ideas of critical exponents, universality and scaling: which form the main %%@
physical ideas in (continuous) phase transitions, and which will prepare us for the %%@
renormalization group.

We again refer to the Ising model; and we start by considering the magnetization \begin{eqnarray}
M =  - \frac{1}{N}\frac{\partial}{\partial B}\left(\frac{\CF}{kT}\right)\, .
    \end{eqnarray}
For a vanishing external %%@
magnetic field $B$, the Ising model in two dimensions or higher can exhibit a {\em %%@
spontaneous magnetization} ($M\neq 0$) below the Curie temperature $T_c$. This spontaneous %%@
magnetization can be understood using the corresponding correlation function $\CG_n$, which %%@
measures the response of a spin at the origin $\sigma_0$ to
 a fluctuation occurring at some site $n$
\be
\CG_n = \langle \sigma_n\sigma_0\rangle\, , %%@
\label{correlationfunction}
\ee
where $\langle\ldots\rangle$ is the statistical average. More specifically, the spontaneous magnetization (and in general, a phase transition) is %%@
described by {\em singularities} associated with the  correlation function $\CG_n$.
For large temperatures ($T>T_c$) the thermal fluctuations are dominant over the spin-spin %%@
interactions and the spins become uncorrelated for large $\vert n\vert$. In the asymptotic %%@
situation, $\vert n\vert\to\infty$, the correlation function falls off exponentially with %%@
$\vert n\vert$
\be
\CG_n \sim \exp\Big(-\vert n\vert/\xi(T)\Big)\, ,
\ee
where $\xi(T)$ is the {\em correlation length}. This correlation length summarizes the %%@
average length-scale on which microscopic quantities' values are correlated. At (and near) %%@
the Curie temperature, the correlation function  can be shown to display a {\em power-law %%@
decay}
\be
\CG_n \sim 1/\vert n\vert^{d-2+\eta}\, ,\label{power1}
\ee
where $d$ is the spatial dimension and $\eta$ is the {\em critical exponent}. The $(d-2)$ %%@
contribution can be deduced from simple dimensional considerations about the physical %%@
quantities of the theory. On the other hand, the critical exponent $\eta$ (which is also %%@
called {\em anomalous}) is the ``non-trivial'' contribution to the correlation function $\CG$. The deduction of this anomalous exponent %%@
$\eta$ is widely considered to be one of  the most important achievements of the renormalization group %%@
(Fisher (1998, pp. 657-659)).

We will now be interested in the behaviour of the physically relevant quantities (mainly, %%@
thermodynamic quantities) near or at the critical point. We start by considering the %%@
behaviour  of the  magnetization $M$ (for a vanishing external field $B$) near the critical %%@
temperature, in terms of the reduced (dimensionless) temperature $\tau\equiv (T-T_c)/T_c$. %%@
Like the correlation function $\CG_n$, the magnetization $M$  is described by a power-law
\be
M \sim \vert\tau\vert^\beta,\, \hspace{0.5cm}{\rm as}\hspace{0.5cm} \tau \to 0^{-}\label{power2}
\ee
where $\beta$ is another critical exponent. Power-law behaviour is characteristic of the %%@
physics near the critical temperature. Other quantities such as the correlation function %%@
$\xi$, the heat capacity $\CC$ (for a vanishing external field) and the isothermal %%@
susceptibility $\chi$, also exhibit power-law behaviour (as $\tau \to 0^\pm$):
\be
\xi \sim\, \vert\tau\vert^{-\nu} , \hspace{1cm} \CC\sim \vert\tau\vert^{-\alpha}\, %%@
,\hspace{1cm} \chi\sim\vert\tau\vert^{-\gamma}\, ,\label{critexponenst}
\ee
where $\nu$, $\alpha$ and $\gamma$ are also critical exponents. And they are found to satisfy %%@
certain algebraic relations:  $\gamma = (2-\eta)\nu$ and $\alpha + 2\beta + \gamma = 2$.

One of the most remarkable aspects of critical phenomena is their {\em universal} nature. %%@
That is: power-law behaviour, and even the numerical values of the critical exponents (and so %%@
their algebraic relations), are not specific to our particular  example of the ferromagnet. %%@
They also describe various physical systems, which are completely different from the %%@
ferromagnet, such as the gas/liquid transition in fluids. So different physical systems with %%@
the same numerical values for the critical exponents are said to belong to the same {\em %%@
universality class}. Understanding the origin of {\em universality}, and classifying various %%@
critical regimes in different universality classes, is another of the remarkable results of %%@
the renormalization group.

Another peculiar property of the physics near criticality is the behaviour of the correlation %%@
length $\xi$. One notices from Eq.(\ref{critexponenst}) that the correlation length $\xi$ goes %%@
to infinity  as the Curie temperature is approached ($\xi\to\infty$ for $T\to T_c$). We %%@
remark that the divergent character of correlation length is related to the masslessness %%@
limit in quantum field theory, since $\xi^{-1}\to 0$.\footnote{The similarities between %%@
statistical mechanics  and quantum field theory go far beyond the behaviour of the %%@
correlation length near criticality. But we cannot here enter this rich field.}

So far, we have discussed the physics near the critical point in terms of the correlation %%@
function $\CG_n$ of Eq.(\ref{correlationfunction}). It is natural to wonder  if one could %%@
understand some of this physics using some more familiar concepts of statistical mechanics, %%@
such as the partition function or the free energy. Indeed, one can: we will now discuss
critical physics using again the free energy $\CF$.

For the Ising model, we have so far considered the free energy as a function of the %%@
temperature $T$ and of a possible external magnetic field $B$: $\CF(T,B)$. But now, %%@
anticipating a little the renormalization group analysis of universality, we envisage the %%@
introduction of extra variables, collectively called $P_i$; (they could be the pressure or %%@
dipole-dipole couplings). So we consider the free energy: $\CF(T,B,P_i)$.

One way of understanding the properties   of the critical exponents in terms of $\CF$ is to %%@
consider the {\em singular} part of the free energy, $f_s$, (more precisely the singular part %%@
of the free energy density, i.e. the free energy per unit of volume): which is obtained by %%@
subtracting  from  the original free energy the analytic part of $\CF$. It will turn out more %%@
convenient to express the singular free energy in terms of a new family of reduced variables: %%@
the reduced temperature $\tau := (T - T_c)/T_c$ (introduced at Eq. \ref{power2} to describe power-law behaviour near
criticality),  $h := \mu B/k_B T $ and $g_i := P_i / k_B T$. (In quantum field theory, the %%@
$g_i$ correspond to the coupling constants, and we will use this terminology in the following %%@
discussion.) We are now ready to describe the critical behaviour of $f_s(\tau, h, g_i)$.

In terms of reduced variables and the critical exponents, the singular free energy obeys the %%@
{\em scaling hypothesis}
\be
f_s(\tau,h,g_i) \approx \vert\tau\vert^{2-\alpha} \Psi %%@
\left(\frac{h}{\vert\tau\vert^{\beta+\gamma}}, \frac{g_i}{\vert\tau\vert^{\phi_i}}\right)\, ; %%@
\label{scalinghypo}
\ee
where $\beta$ and $\gamma$ are the critical exponents associated respectively to the %%@
magnetization $M$ and the heat capacity $\CC$, and  $\Psi$ is known as the {\em scaling %%@
function}. When appropriately normalised, $\Psi$ is also universal. From the perspective of the renormalization group, the crucial aspect of %%@
Eq.(\ref{scalinghypo})  is the appearance of a set of new critical exponents $\{\phi_i\}$ %%@
accompanying the coupling constants $g_i$.

Before discussing these new critical exponents $\phi_i$, we should note the qualitative %%@
difference between the descriptions of critical physics by the singular free energy $f_s$ %%@
and by the power-laws involving the correlation function $\CG_n$. Namely, the critical %%@
exponents $\phi_i$ in the scaling hypothesis Eq.(\ref{scalinghypo}) are absent from the power-laws in Eq.s (\ref{power1}), (\ref{power2}) and (\ref{critexponenst}). This prompts the
question: do the descriptions simply contradict each other, or can they be %%@
reconciled? As one might guess, the answer is (fortunately!) the latter. A full
understanding of this situation involves the renormalisation group. But one can get %%@
considerable insight into  the situation by looking at the asymptotic behaviour (when %%@
$\tau\to 0$) of the argument $g_i/\vert\tau\vert^{\phi_i}$; as follows.

The signs of the exponents $\phi_i$  will fix the asymptotic behavior of the argument %%@
$g_i/\vert\tau\vert^{\phi_i}$; (their specific numerical values will be irrelevant). Thus %%@
consider a situation with only one exponent $\phi_i = \phi$, whose sign is negative. Then the %%@
argument $g_i/\vert\tau\vert^{\phi_i}$ will become  very small near the critical
temperature: $g_i/\vert\tau\vert^{\phi_i}\to 0$. From Eq.(\ref{scalinghypo}), the critical %%@
behaviour of the scaling function $\Psi$, for negative $\phi$, will be  independent of the %%@
couplings $g_i$, and  one can write
\be
 \Psi \left(\frac{h}{\vert\tau\vert^{\beta+\gamma}}, \frac{g}{\vert\tau\vert^\phi}\right) \to  %%@
\Psi \left(\frac{h}{\vert\tau\vert^{\beta+\gamma}}, 0\right)\, . \label{scaleg0}
\ee
From Eq.(\ref{scaleg0}), one concludes that the (singular) free energy at the critical %%@
temperature  is completely independent of the argument  $g_i/\vert\tau\vert^{\phi_i}$. As a %%@
consequence of this, one can recover, starting from the scaling function $\Psi$ in %%@
Eq.(\ref{scaleg0}), the power-law equations in Eq.s (\ref{power1}), (\ref{power2}) and %%@
(\ref{critexponenst}); (with their independence of the new exponents $\phi_i$).

The moral of this story is that very different physical systems, described by different set %%@
of couplings $g_i$---but for which all the corresponding exponents $\phi_i$ have a negative %%@
sign---are described by the same scaling function $\Psi %%@
\left(h/\vert\tau\vert^{\beta+\gamma}, 0\right)$. So these systems will have the same %%@
physical properties at the critical point. Thus the contributions of the couplings $g_i$ (or %%@
their associated quantities) are called {\em irrelevant}. This again expresses the idea of %%@
universality.

The description in terms of the renormalisation group is to envisage a large (in general  %%@
infinite-dimensional) space of Hamiltonians, with coordinates $(\tau, h, g_i)$. Then
a set of Hamiltonians that differ from each other only by irrelevant couplings will exhibit %%@
the same behaviour at the critical point. We will see in Section \ref{WRG} that this is a %%@
fixed point of a flow on the space. One  also envisages that the space of Hamiltonians might %%@
contain several, even many, fixed points: these different fixed points will describe %%@
different universality classes, with different critical exponents and scaling functions.

We now have a good understanding of the case where the exponent $\phi$ is negative. Let us %%@
now turn to the other case where the sign of $\phi$ is positive. In this case, the argument %%@
$g/\vert\tau\vert^\phi$ of the scaling function grows larger and larger near the critical %%@
point ($\tau\to 0$); and so this contribution is  called {\em relevant}. This growth of the %%@
argument engenders two possible situations:\\
\indent (i): the fixed point is ``destroyed'' by the relevant contributions; for example, %%@
this happens in the Ising model when one takes into account the %%@
magnetic field $B$ (from Eq.(\ref{scalinghypo}) one notices that the exponent associated with the magnetic field is $\beta+\gamma$. This exponent is positive, and therefore this situation corresponds to a relevant perturbation).\\
\indent  (ii): if there are multiple fixed points, the effect of the relevant contributions %%@
is to send the system from one fixed point (universality class) to a different one (a %%@
different universality class). One may realize this situation in the case of a ferromagnet by considering long-range magnetic dipole-dipole interaction. In Section \ref{crossover}, we will appeal to this sort of behaviour, %%@
called {\em cross-over}, to discuss phase transitions in finite systems and to illustrate our claim (2:Before).

\subsection{The Wilsonian approach to the renormalization group}\label{WRG}
In the previous Section, we have already glimpsed some aspects of the renormalization group. We said that it envisages a space of Hamiltonians, on which is defined a flow, and that the critical point is a fixed point of this flow (in general one of several such points). We also said that understanding the ideas of anomalous exponent, universality
classes, and irrelevant and relevant quantities were major achievements of the renormalisation %%@
group perspective. In this Section, we will give more details, albeit at a conceptual level.\footnote{We happily follow the widespread courtesy of dubbing the approach `Wilsonian', in honour of Kenneth Wilson; (Wilson and Kogut (1974) provides an excellent introduction to Wilson's ideas). But we emphasize that several other {\em maestros} contributed mightily to its development: within statistical mechanics, two obvious examples are Michael Fisher and Leo Kadanoff.}

There will be two leading ideas. First: the flow is defined by reiterating a procedure of coarse-graining that averages over the details of the physics at short distances and so defines an effective Hamiltonian. Then the fact that at the critical point, the correlation length diverges means that there, the system ``looks the same at all length scales''---which is represented mathematically by the critical point being a fixed point of the flow. Second: we need to allow that under coarse-graining, the form of the Hamiltonian changes, maybe radically. So we envisage the effective Hamiltonians to include all possible terms, so that the space of Hamiltonians, which is coordinatized by the various interaction strengths between spins/sites/particles and couplings to external fields (and typically, also temperature) is in general infinite-dimensional; (though in any single effective Hamiltonian, many couplings might be zero).

We begin the exposition with the idea of coarse-graining. Recall that at the critical point, the
correlation length $\xi$ diverges:  $\xi \to\infty$. So near it, $\xi\gg a$, where $a$ is the lattice spacing; and the physics of the system
will be governed by modes whose wavelengths are much larger than $a$. The strategy of the %%@
renormalisation group is, accordingly, to integrate out the short wavelengths.\footnote{Similarly in %%@
quantum field theory: we integrate out the high-frequency modes and the system is governed by %%@
the low frequency modes for which $\omega < a^{-1}$, where $\Lambda \equiv a^{-1}$ is the %%@
cutoff.} As a result of this integration, one is left with an {\em effective} Hamiltonian, %%@
which is defined only in terms of the long wavelength modes, but which has the same critical %%@
properties as the original system.

For example, one way of integrating out short wavelengths is Kadanoff's block-spin approach. %%@
In the Ising model, the Kadanoff approach consists of coarse-graining by combining several %%@
neigbouring spins into a block, each block then being assigned a single new spin variable %%@
(by, say, majority vote). The crucial aspect of this procedure is that the new block-spin %%@
system is also an Ising model, described by an effective Hamiltonian, with an effective %%@
temperature and an effective magnetic field. In more detail: the idea is to reduce the number %%@
of degrees of freedom of the original system (``integrate out''), by dividing the original %%@
lattice into blocks of $b^d$ spins.\footnote{The parameter $b$ is large but less than the %%@
ratio $\xi/a$. Since $\xi$ diverges at the critical point, in effect $b$ can be chosen %%@
arbitrarily. As before, $d$ is the spatial dimension.} This operation transforms a lattice %%@
system with $N$ spins into a new system with fewer degrees of freedom, $N'=N/b^d$ represented %%@
by new (rescaled) spin variables $\sigma'_{n'}$. The new lattice $N'$ will reproduce the %%@
large-scale properties of the original lattice $N$ if one rescales all the space (lattice) %%@
coordinates by: $x' = x/b .$

One can state this more formally as a three-step procedure, applied to the Hamiltonian. It is convenient to begin by  defining the reduced hamiltonian
\be
H [\{\sigma_n\}; {K}] = -\H[\{\sigma_n\}; {K}]/k T \equiv - \beta {\cal H} \,
\ee
where $\{\sigma_n\}$ represents configurations, and $K$ encodes the coupling $J$ and $\mu$ in Eq. \ref{IsiHamn}. We will soon generalize $K$ to include other couplings; but for the moment, our notation can suppress it. With this reduced Hamiltonian, one then takes three steps.\\
 \indent (i): One divides the set of $N$ spin %%@
variables $\sigma_n $ into two families; first, the long wavelength modes $\sigma_n^L$, %%@
consisting of $N' = N/b^d$ spins; second, the short wavelength modes $\sigma_n^S$ consisting %%@
of the remaining $N-N'$ variables.\\
\indent (ii): Then one takes the sum over
the short wavelength modes. For the partition function, this means writing
\be
\sum_{\sigma^L, \sigma^S} \exp(H[\sigma^L,\, \sigma^S]) = %%@
\sum_{\sigma^L}\exp(H_{\rm{eff}}[\sigma^L])\, ,
\ee
where
\be
\exp(H_{\rm{eff}}[\sigma^L]) = \sum_{\sigma^S} \exp(H[\sigma^L,\, \sigma^S])\, .
\ee
Thus one is left with an {\em effective} Hamiltonian $H_{\rm{eff}}[\sigma^L]$, involving only
the spins $\sigma_n^L$. The crucial aspect of the Wilsonian approach is that %%@
$H_{\rm{eff}}[\sigma^L]$  can include all  possible terms (and couplings), in general an %%@
infinite  number of them, thus defining an infinite-dimensional space of Hamiltonians.\\
\indent (iii): Following the spatial rescaling $x' = x/b$, one can obtain %%@
renormalised (i.e. rescaled) spin variables $\sigma'$ and their corresponding renormalised %%@
Hamiltonian $H'[\sigma']\equiv H_{\rm{eff}}[\sigma^L]$. We write this as
\be
H'[\sigma'] =\CR_b \{H[\sigma]\}\, ,\label{transfhamiltonian}
\ee
where $\CR_b$ is called the renormalization group operator.

We now envisage {\em iterating} the transformation in Eq.(\ref{transfhamiltonian}), thus obtaining a flow: a sequence of renormalized Hamiltonians
\be
H^{(\ell)} = \CR_b [H^{(\ell-1)}]=(\CR_b)^l [H]\, ,
\ee
where $H^0\equiv H$ and $H^1\equiv H'$. (So the powers of $\CR_b$ form a semigroup; but the more accurate phrase `renormalization semigroup' has never caught on!). More generally, we sometimes envisage continuous transformations of the Hamiltonians. Then we may express the flow %%@
equations  by a differential equation
\be
\frac{d}{d\ell}H =\CB[H]\, .
\ee
This form is familiar in quantum field theory, where $\CB$ is the beta-function.

The fixed points of this flow (i.e. Hamiltonians left unchanged by the
renormalisation group transformations) are then defined by the equations:
\be
\CR_b[H^\ast] = H^\ast\, , \hspace{0.5cm} or \hspace{0.5cm}  \CB[H^\ast] =0\, .
\ee

At this point, we recall that the effective Hamiltonians can be coordinatized by their set of coupling %%@
constants $K$. Thus the renormalization group operator $\CR_b$ relates the renormalized couplings $K'$ to the original %%@
couplings $K$ via the {\em recursion relations}
\be
K' = R_b K \label{recrel}\, .
\ee
So the fixed points correspond to couplings that are left unchanged %%@
by the renormalisation transformation:
\be
K^\ast = R_b   K^\ast\, .\label{fixcoup}\,
\ee
One can understand much of the physics near critical points in terms of the transformations of the couplings. As an example, we will close this Section by using Eq.(\ref{fixcoup}) to characterize Section \ref{critexp}'s notions of  relevant and irrelevant behaviours near the critical point.

Within a neighbourhood of the fixed point, we linearise the couplings: $K_n = K_n^\ast + \delta K$, and  $K'_n = K_n^\ast + \delta K'$. Applying the recursion relation in Eq.(\ref{recrel}), we obtain 
\be
K'_n = K^\ast_n + \sum_m T_{mn} \delta K_m\, ,
\ee
where $T_{mn} = \left.\frac{\partial K'_n}{\partial K_m}\right |_{K_m =K_m^\ast}$. We  shall assume %%@
for simplicity  that the matrix $T$ is  symmetric, and examine its eigenvalues in terms of the length scale $b$ raised to a certain power $y_i$
\be
\sum_m\phi_n^i T_{nm} = b^{y_i}\phi^i_m\, .
\ee
The sign of the exponents $y_i$ gives the stability properties of the fixed point. This can %%@
be seen by defining scaling variables $u_i := \sum_m\phi^i_m\delta K_m$, which are linear %%@
combinations of deviations away from the fixed point. Their significance comes from the fact %%@
that they transform linearly under $R_b$ near the fixed point:
\be
u_i' = b^{y_i} u_i\, .
\ee
So we may classify the different properties of the flow behaviour according to the signs of the %%@
eigenvalues.
\begin{itemize}

\item $y_i < 0$: $u_i$ decreases under the renormalisation group transformations, the system flows towards the 
fixed point. We say we have an {\em irrelevant} eigenvalue. And similarly, we speak of an irrelevant direction and scaling variable.

\item $y_i > 0$:  $u_i$ increases, the system 
flows away from the fixed point. We say that the  eigenvalue is {\em relevant}; and similarly, the direction 
and scaling variable is relevant.

\item $y_i =0$: This situation is  more complicated: we cannot tell the behaviour of the flow from the linearized 
equations. This situations is called {\em marginal}.
\end{itemize}

We end this Section with a brutal summary of the idea of the renormalisation group; as follows. One defines a space $X$ coordinatized by the parameter values that define the microscopic Hamiltonian. One defines a transformation $R$ on $X$ designed to preserve the large-scale %%@
physics of the system. Typically, $R$ is a coarse-graining, defined by local collective 
variables that take some sort of majority vote about the local quantities' values, followed %%@
by a rescaling, so that the resulting system can be assigned to a point in %%@
$X$. This assignment of the resulting system to a point within X enables one to %%@
consider iterating $R$, so that we get a flow on $X$. Critical points where $\xi$ %%@
diverges will be among the fixed points of this flow. For the fact that $\xi$ diverges means %%@
that the system ``looks the same at all length scales''.

%%%%%%%%%%%%%%%%%%%%%%%%%%%%%%%%%%%%%%%%%%%%%%
\section{The two claims illustrated by phase transitions}\label{emergephtrn}
%%%%%%%%%%%%%%%%%%%%%%%%%%%%%%%%%%%%%%%%%%%%%%%

In this Section, we first summarize how phase transitions illustrate Section \ref{peace}'s two claims, (1:Deduce) and (2:Before), and then endorse a  %%@
proposal of Mainwood's about emergence before the limit: i.e. about how to think of phase %%@
transitions in finite-$N$ systems (Section \ref{phtrnwithwithoutpaul}). Then, using Sections \ref{approach} and \ref{WRG}, we will report in Section \ref{crossover} a remarkable class of phenomena %%@
associated with phase transitions, viz. cross-over phenomena; (these are also emphasized by Mainwood).
These phenomena make emergence before the limit even more vivid than it was in our previous %%@
examples; for they show how an emergent phenomenon can be first gained, then {\em lost}, as %%@
we approach a phase transition. Besides, this will illustrate not only our claim (2:Before), but also another moral argued for by Butterfield (2011): that considerations about very large $N$ show that the $N\to\infty$ limit of our models is unrealistic.

%%%%%%%%%%%%%%%%%%%%%%%%%%%%%%%%%%%%%%%%%%
\subsection{Emergence at the limit and before it}\label{phtrnwithwithoutpaul}
%%%%%%%%%%%%%%%%%%%%%%%%%%%%%%%%%%%%%%%%%%%

Our main claims are (1:Deduce) and (2:Before). Applied to phase
transitions, they say, roughly speaking:\\
\indent (1:Deduce): Some of the emergent behaviours shown in phase
transitions are, when understood (as in thermodynamics) in terms of non-analyticities, %%@
rigorously deducible within a statistical mechanical theory that takes an appropriate version %%@
of the $N \raw \infty$ limit.\\
\indent (2:Before): But these behaviours can also be understood more weakly; (no doubt, this %%@
is in part a matter of understanding them phenomenologically). And thus understood, they %%@
occur before the limit, i.e. in finite-$N$ systems.

Here we admit that the phrases `some of the emergent', `appropriate version' and `can be %%@
understood more weakly' are vague: hence our saying `roughly speaking'. To overcome this vagueness one would have to define, for (1:Deduce): (a) a handful of novel and robust 
behaviours shown in phase transitions (a handful of $T_b$s), and (b) a corresponding handful %%@
of statistical mechanical theories $T_t$ in which the behaviours are rigorously deducible {\em if}
one takes an appropriate  version of the thermodynamic limit, $N \raw \infty$. Similarly, as regards overcoming the vagueness for (2:Before). 

Of course, there is no space here to give such definitions. But we submit that they could be given, for a wide class of emergent behaviours; and 
that Section \ref{phtrexpound} gives  evidence for this. In particular, we mention that Lee-Yang theory (Section \ref{YLKMS}) provides an example of
 (1:Deduce); and the Ising model (Section \ref{Ising}) exemplifies (2:Before).

We turn to endorsing a proposal of Mainwood's (2006, Section %%@
4.4.1, p. 238; 2006a, Section 4.1), which fits well with the swings-and-roundabouts flavour of        %%@
combining (1:Deduce) with %%@
(2:Before). Mainwood's topic is the recent philosophical %%@
debate about phase transitions in finite systems, especially as formulated by Callender
(2001, p. 549) in terms of four jointly contradictory propositions. Mainwood first gives a very judicious survey of the pros and cons of denying %%@
each of the four propositions. Then he uses his conclusions to argue for a proposal that %%@
evidently reconciles: (a) statistical mechanics' use of the thermodynamic limit to describe %%@
phase transitions; and  (b) our saying that phase transitions %%@
occur in the finite system.
That is, to take a stock example: Mainwood's proposal secures that a kettle of water, though %%@
a finite system, can boil!

Mainwood's proposal is attractively simple. It is that for a system with $N$ degrees of %%@
freedom, with a free energy $F_N$ that has a well-defined thermodynamic limit, $F_N \raw %%@
F_{\infty}$, we should just say:\\
\indent\indent  phase transitions occur in the finite system iff $F_{\infty}$ has %%@
non-analyticities.\\
And if we wish, we can add requirements that avoid our having to say that small systems (e.g. %%@
a lattice of four Ising spins laid out in a square) undergo phase transitions. Namely: we  %%@
can add to the above right-hand side conditions along the following lines: {\em and} if $N$ %%@
is large enough, or the gradient of $F_N$ is steep enough etc. Agreed, `large enough' %%@
etc. are vague words. But Mainwood thinks, and we agree, that the consequent vagueness about whether a phase 
transition occurs is acceptable: after all, `boil' etc. are vague.

%%%%%%%%%%%%%%%%%%%%%%%%%%%%%%%%%%%%%%%%%%%%%%
\subsection{Cross-over: gaining and losing emergence at finite $N$}\label{crossover}
%%%%%%%%%%%%%%%%%%%%%%%%%%%%%%%%%%%%%%%%%%%%%%%%

We end by describing {\em cross-over phenomena}. We again follow Mainwood, who uses it (2006, %%@
Sections 4.4.2-3, pp. 242-247; 2006a, Section 4.2) to illustrate and defend his proposal for %%@
phase transitions in finite systems. We concur with that use of it. But our own aims are %%@
rather different. The main idea  will be that cross-over phenomena yield striking %%@
illustrations of ``oscillations'' between (2:Before) and behaviour that shows that for very large $N$ our models become \emph{unrealistic}. (Butterfield (2011, Section 2) discussed this sort of behaviour under the label (4:Unreal).) That is: a system can be:\\
\indent (i) first,  manipulated so as  to illustrate (2:Before), i.e. an emergent behaviour %%@
at finite $N$; and \\
\indent (ii) then manipulated so as  to lose this behaviour, i.e. to illustrate the model becoming {\em %%@
unrealistic}; with the manipulation corresponding to higher values %%@
of $N$; and  \\
\indent  (iii) then manipulated so as  to either (a) enter a regime illustrating some other %%@
emergent behaviour, or (b) revert to the first emergent behaviour; so that either (a) or (b) %%@
illustrate (2:Before) again.\\
In short: cross-over will illustrate the swings-and-roundabouts combination of (2:Before) and our models becoming unrealistic for very large $N$.\footnote{Cross-over also illustrates another point about (2:Before), which is also seen in other examples in Butterfield (2011): viz. how the emergent behaviour at large but finite $N$, can be ``lost'' if one alters certain features of the situation.}

Cross-over phenomena  occur near a 
critical point,  where the  correlation length $\xi$ diverges. The idea is that at first the system appears to show behaviour characteristic of one universality class around the point; but as it is brought 
even closer to the point, the behaviour rapidly changes to that characteristic of another point or universality class.  (For details, cf. e.g. Yeomans (1992, p. 112), Cardy (1997, pp. 61, 69-72), Chaikin and Lubensky (2000, pp. 216, 270-3); Hadzibabic et al. (2006) is a recent example of experiments.)

For example, let us consider {\em finite-size %%@
cross-over}. This occurs when the ratio of $\xi$ to the system's size determines the fixed %%@
point towards which the renormalization group flow sends the system. When $\xi$ is small compared to the size of 
the system, though very large on a microscopic length-scale, the system flows towards a %%@
certain fixed point representing a phase transition; and so exemplifies a certain %%@
universality class. Or to put it more prosaically: coarser and coarser (and suitably %%@
rescaled) descriptions of the system are more and more like descriptions of a phase %%@
transition. So in the jargon of our claims: the system illustrates (2:Before). But as $\xi$ %%@
grows even larger, and becomes comparable with the system size, the flow crosses over and %%@
moves away---in general, eventually, towards a different fixed point. In our jargon: the %%@
system runs up against the model's being  unrealistic, and goes over to another %%@
universality class---eventually to another behaviour again illustrating (2:Before).

 Of course, the correlation length will only approach a system's physical size when the %%@
system has been brought very close indeed to the phase transition, well within the usual %%@
experimental error. That is: until we enter the cross-over regime, experimental data about %%@
quantities such as the gradient of the free energy will strongly suggest non-analyticities, %%@
such as a sharp corner or an infinite peak. Or in other words: until we enter this regime, %%@
the behaviour will be as if the system is infinite in extent. But once we enter this regime, %%@
and the cross-over occurs, the appearance of non-analyticities goes away: peaks become tall %%@
and narrow---but finitely high. Again, in our jargon, we have: (2:Before) followed by the model's being revealed as unrealistic.

Finally, we note that a similar discussion would apply to other kinds of cross-over, such as {\em %%@
dimensional cross-over}. For example, this occurs when the behaviour of a thin film crosses %%@
over from a universality class typical of three-dimensional systems to one for %%@
two-dimensional systems, as $\xi$ becomes comparable with the film's thickness.

\section{Conclusion}\label{concl}
Needless to say, we have only scratched the surface of our chosen field! Thermodynamics and statistical mechanics are vast sciences: we believe that they contain many other examples illustrating our claims, (1:Deduce) and (2:Before). For instance, to stick to the $N \raw \infty$
limit of quantum theory: there is KMS states' description of thermodynamic phases (Section \ref{YLKMS}). 
Showing the claims in many such examples would indeed be strong testimony to the 
reconciliation of emergence and reduction. Work for another day!

{\em Acknowledgements}:--- It is a pleasure to thank the organizers of the conference, Frontiers of Fundamental Physics 11, and colleagues there, for a most pleasant and stimulating meeting. We are also very grateful to the editors---not least for their patience!  This publication was made possible through the support of a grant from Templeton World Charity Foundation. The opinions expressed in this publication are those of the authors and do not necessarily reflect the views of Templeton World Charity Foundation.

\section{References}

Anderson, P. (1972), `More is different', {\em Science} {\bf 177}, pp. 393-396; reprinted in %%@
Bedau and Humphreys (2008).

Bangu, S. (2009), `Understanding thermodynamic singularities: phase transitions, data and %%@
phenomena', {\em Philosophy of Science} {\bf 76}, pp. 488-505.

Batterman, R. (2005), `Critical phenomena and breaking drops: infinite idealizations in %%@
physics', {\em Studies in History and Philosophy of Modern Physics} {\bf 36B}, pp. 225-244.

Blythe, R. A. and Evans, M.R. (2003), `The Lee-yang theory of equilibrium and nonequilibrium phase transition', {\em The Brazilian Journal of Physics} {\bf 33}, pp. 464-475.

Butterfield, J. (2011), `Less is different: emergence and reduction reconciled', %%@
forthcoming in  {\em Foundations of Physics}. 
Available at http://philsci-archive.pitt.edu/8355/

Butterfield, J. (2011a), `Emergence, reduction and supervenience: a varied landscape', %%@
forthcoming in {\em Foundations of Physics}. 
Available at http://philsci-archive.pitt.edu/5549/

Callender, C. (2001), `Taking thermodynamics too seriously', {\em Studies in History and %%@
Philosophy of Modern Physics} {\bf 32B}, pp. 539-554.

Cardy, J. (1997), {\em Scaling and Renormalization in Statistical Physics}, Cambridge Lecture %%@
Notes in Physics, volume 5; Cambridge University Press.

Chaikin, P. and Lubensky, T. (2000), {\em Principles of Condensed Matter Physics}, Cambridge %%@
University Press.

Emch, G. (1972), {\em Algebraic Methods in Statistical Mechanics and Quantum Field Theory}, %%@
John Wiley.

Emch, G. (2006), `Quantum Statistical Physics', in {\em Philosophy of Physics, Part B}, a %%@
volume of {\em The Handbook of the Philosophy of Science}, ed. J. Butterfield and J. Earman, %%@
North Holland, pp. 1075-1182.

Emch, G., and Liu, C. (2002), {\em The Logic of Thermo-statistical Physics}, Springer-Verlag

Fisher, M. E. (1998), `Renormalization group theory: Its basis and formulation in statistical physics', {\em Rev. Mod. Phys} {\bf 70}, pp 653-681.

Gross, D. (2001), {\em Microcanonocal Thermodynamics; phase transitions in small systems}, %%@
World Scientific.

Hadzibabic, Z. et al. (2006), `Berezinskii-Kosterlitz-Thouless crossover in a trapped atomic %%@
gas', {\em Nature} {\bf 441}, 29 June 2006, pp. 1118-1121.

Kadanoff, L.  (2009), `More is the same: phase transitions and mean field theories', {\em %%@
Journal of Statistical Physics} {\bf 137}, pp. 777-797; available at %%@
http://arxiv.org/abs/0906.0653

Kadanoff, L. (2010), `Theories of matter: infinities and renormalization', available at %%@
http://arxiv.org/abs/1002.2985; and at http://jfi.uchicago.edu/~leop/AboutPapers/Trans2.pdf.

Kadanoff, L. (2010a), `Relating theories via renormalization', available at\\  %%@
http://jfi.uchicago.edu/~leop/AboutPapers/RenormalizationV4.0.pdf

Lavis, D. and Bell, G. (1999), {\em Statistical Mechanics of Lattice Systems 1; Closed Forms %%@
and Exact Solutions}, Springer: second and enlarged edition.

Lavis, D. and Bell, G. (1999a), {\em Statistical Mechanics of Lattice Systems 2; Exact, %%@
Series and Renormalization Group Methods}, Springer.

Le Bellac, M. (1991), {\em Quantum and Statistical Field Theory}, (translated by G. Barton) %%@
Oxford University Press.

Lee, T. D. and Yang, C. N. (1952), `Statistical theory of equations of state and phase transitions. II. Lattice gas and Ising model', {\em Physical Review} {\bf 87}, pp. 410-419.

Liu, C. (2001), `Infinite systems in SM explanation: thermodynamic limit, renormalization %%@
(semi)-groups and irreversibility' {\em Philosophy of Science} {\bf 68} (Proceedings), pp. %%@
S325-S344.

Liu, C. and Emch, G. (2005), `Explaining quantum spontaneous symmetry breaking', {\em Studies %%@
in History and Philosophy of Modern Physics} {\bf 36}, pp. 137-164.

Mainwood, P. (2006), {\em Is More Different? Emergent Properties in Physics}, D.Phil.
dissertation, Oxford University. Available at: http://philsci-archive.pitt.edu/8339/

Mainwood, P. (2006a), `Phase transitions in finite systems', unpublished MS; (corresponds to %%@
Chapter 4 of Mainwood (2006). Available at: http://philsci-archive.pitt.edu/8340/

Nagel, E. (1961), {\em The Structure of Science: problems in the Logic of Scientific %%@
Explanation}, Harcourt.

Nagel, E. (1979), `Issues in the logic of reductive explanations', in his {\em Teleology Re-
visited and other essays in the Philosophy and History of Science}, Columbia University
Press.

Ruelle, D. (1969), {\em Statistical Mechanics: Rigorous Results}, W.A. Benjamin.

Sewell, G. (1986), {\em Quantum Theory of Collective Phenomena}, Oxford University Press.

Sewell, G. (2002), {\em Quantum Mechanics and its Emergent Microphysics}, Princeton %%@
University Press.

Thompson, C. (1972), {\em Mathematical Statistical Mechanics}, Princeton University Press.

Wilson, K. and Kogut, J. B. (1974), `The renormalization group and the $\epsilon$ expansion'. {\em Physics Reports} {\bf12}, pp. 75-200.

Yeomans, J. (1992), {\em Statistical Mechanics of Phase Transitions}, Oxford University %%@
Press.

\end{document}